# Partial Oxidation of Methane on a Nickel Catalyst: Kinetic Monte-Carlo Simulation Study

Sirawit Pruksawan [a], Boonyarach Kitiyanan [a,*], Robert M. Ziff [b]

[a] The Petroleum and Petrochemical College, Chulalongkorn University, Bangkok 10330, Thailand
[b] Department of Chemical Engineering, University of Michigan, Ann Arbor, MI 48109, USA
*e-mail: boonyarach.k@chula.ac.th

**Abstract**

Kinetic Monte-Carlo simulation is applied to study the partial oxidation of methane over a nickel catalyst. Based on the Langmuir-Hinshelwood mechanism, the kinetic behavior of this reaction is analyzed and the results are compared with previous experiments. This system exhibits kinetic phase transitions between reactive regions with sustained reaction and poisoned regions without reaction. The fractional coverages of the adsorbed species and the production rates of $H_2$, $CO$, $H_2O$, and $CO_2$ are evaluated at steady state as functions of feed concentration of the methane and oxygen, and reaction temperature. The influence of lattice coordination number, diffusion, and impurities on the surface is also investigated. The simulation results are in good agreement with the experimental studies where such results are available. It is observed that when the lattice coordination number is increased to eight, the width of the reactive region increases significantly. Moreover, the phase transition becomes continuous. The diffusion of adsorbed O and H on the surface plays a measurable role in the reaction, increasing the maximum production rates as the diffusion rate increases. In systems with impurities, the production rates are greatly reduced and the phase transition is also changed from being abrupt to continuous.

Keywords: Kinetic Monte-Carlo simulation, Partial oxidation, Kinetics, Diffusion, Impurities

## 1. Introduction

The catalytic conversion of methane is one of the most attractive areas of research in both academia and industry. Steam reforming has been used in industry to produce hydrogen or syngas from methane. Nevertheless, steam reforming is an energy-intensive process. The catalytic partial oxidation (CPOx) process is an attractive alternative because it avoids the need of large amounts of energy and requires smaller reactors due to the faster oxidation reaction. Another advantage of the CPOx process is that the $H_2$-to-CO ratio in syngas products is 2:1, which allows a straightforward syngas utilization for methanol or Fischer–Tropsch synthesis. The CPOx of methane has been extensively studied by various researchers (Hu and Ruckenstein, 1996; Tsipouriari and Verykios, 1998; Smith and Shekhawat, 2011; Al-Sayari, 2013). Even though several catalytic systems have been reported in the literature to be active for this reaction, nickel catalysts are generally believed to be the most promising candidate owing to their moderate cost and good catalytic performance compared to that of noble metals (Al-Sayari, 2013). In spite of the large quantity of experimental studies devoted to the partial oxidation of methane on nickel catalysts, the mechanisms and kinetic studies in this system are still unclear. Theoretical studies have not clarified this issue. Moreover, the published experimental results are contradictory (Smith and Shekhawat, 2011). Therefore, further studies on the kinetic behavior are expected to provide a better understanding to the reaction mechanism.

Kinetic Monte-Carlo (KMC) is an effective tool for investigating surface reaction mechanisms and predicting complex kinetic behaviors. KMC simulation can be viewed as a numerical method to solve the Master Equation that describes the evolution of the catalyst's surface and the adsorbates. The rate constants in this equation are obtained from kinetic studies carried out on the selected catalyst. KMC simulation allows one to visualize the state of surface. One of the simplest kinetic models for the oxidation reaction of CO on a catalytic surface, introduced by Ziff, Gulari and Barshad (ZGB) (Ziff et al., 1986), has provided a source of continual investigation for studies of complex phenomena (Fichthorn et al., 1989; Fichthorn and Weinberg, 1991; Evans, 1993; Tomé and Dickman, 1993; Loscar and Albano, 2003; Machado et al., 2005). Other lattice models have been introduced to simulate different catalytic surface reaction, including CPOx of methane. The CPOx of methane was extensively studied for the oscillatory behavior of the dynamics by a model (Lashina et al., 2012; Ren et al., 2008; Ren and Guoa, 2008). The present work provides information on the kinetic behavior in which the reaction is conducted under steady-state conditions.

The purpose of this work is to develop the KMC simulation for the partial oxidation of methane on a nickel catalyst based on the Langmuir-Hinshelwood (LH) mechanism combined with the formation and removal of nickel oxide under isothermal conditions. The kinetic behavior at steady state is analyzed in different operating conditions and the results are compared with previous experiments. The effects of diffusion of adsorbed species, lattice coordination number and inactive impurities on the surface are also evaluated.

## 2. Model and simulation procedure

### 2.1 Model

**Table 1.** Elementary reactions and corresponding kinetic parameters used in this work

| Steps (i) | Elementary reactions | $A_i$ or $k^0_i$ ($s^{-1}$ or $s^{-1} Pa^{-1}$) | $Ea_i$ (kcal/mol) | References |
|---|---|---|---|---|
| 1 | $CH_4(g) + [*] \rightarrow [CH_4*]$ | 0.0045 ($k^0_i$) | 0 | Ren et al., 2008 |
| 2 | $[CH_4*] \rightarrow CH_4(g) + [*]$ | $1.0 \times 10^4$ | 7.9 | Ren et al., 2008 |
| 3[†] | $[CH_4*] + [*] \rightarrow [CH_3*] + [H*]$ | $1.3 \times 10^8$ | 13.8 | Hei et al., 1998 |
| 4[†] | $[CH_3*] + [*] \rightarrow [CH_2*] + [H*]$ | $1.0 \times 10^{13}$ | 27.6 | Chen et al., 2001a |
| 5[†] | $[CH_2*] + [*] \rightarrow [CH*] + [H*]$ | $1.0 \times 10^{13}$ | 23.2 | Chen et al., 2001a |
| 6[†] | $[CH*] + [*] \rightarrow [C*] + [H*]$ | $1.0 \times 10^{13}$ | 4.5 | Chen et al., 2001b |
| 7 | $O_2(g) + 2[*] \rightarrow 2[O*]$ | 0.011 ($k^0_i$) | 0 | Ren et al., 2008 |
| 8 | $2[O*] \rightarrow O_2(g) + 2[*]$ | $1.0 \times 10^{11}$ | 44.6 | Ren et al., 2008 |
| 9 | $[C*] + [O*] \rightarrow [CO*] + [*]$ | $1.0 \times 10^{12}$ | 35 | Ren et al., 2008 |
| 10 | $[CO*] \rightarrow CO(g) + [*]$ | $1.0 \times 10^{10}$ | 27.85 | Ren et al., 2008 |
| 11[†] | $2[H*] \rightarrow H_2(g) + 2[*]$ | $3.1 \times 10^{12}$ | 23.3 | Chen et al., 2001b |
| 12 | $[CO*] + [O*] \rightarrow CO_2(g) + 2[*]$ | $5.0 \times 10^6$ | 15.2 | Ren et al., 2008 |
| 13 | $[H*] + [O*] \rightarrow [OH*] + [*]$ | $1.0 \times 10^7$ | 19.6 | Li and Xiang, 2000 |
| 14[†] | $[H*] + [OH*] \rightarrow H_2O(g) + 2[*]$ | $3.1 \times 10^{11}$ | 7.7 | Chen et al., 2001a |
| 15 | $[O*] \rightarrow [Ox]$ | $1.0 \times 10^5$ | 15.7 | Ren et al., 2008 |
| 16 | $[C*] + [Ox] \rightarrow [CO*] + [*]$ | $1.0 \times 10^7$ | 26.24 | Ren et al., 2008 |
| 17 | $[CO*] + [Ox] \rightarrow CO_2(g) + 2[*]$ | $1.0 \times 10^5$ | 17 | Ren et al., 2008 |
| 18 | $[H*] + [Ox] \rightarrow [OH*] + [*]$ | $5.2 \times 10^3$ | 11 | Ren et al., 2008 |

**Note.** [X*] = adsorbed X species.
[†] assumed instantaneous

Our kinetic model was developed for the CPOx on a nickel catalyst based on the reaction mechanism of Lashina et al. (2012). According to these authors, the mechanism of CPOx can be described by an 18-step elementary reaction (Table 1). The steps 1–14 describe the CPOx of methane on the surface of nickel metal (denoted by *) with the formation of CO, $H_2$, $CO_2$ and $H_2O$. The formation of nickel oxide (Ox) in step 15 leads directly or indirectly to the formation of CO, $CO_2$ and $H_2O$ with the participation of the oxidized form of nickel in steps 16-18 (Shen et al., 1998).

The estimation of the kinetic parameters of elementary reactions was obtained from the literature as summarized in Table 1. Here, steps 3, 4, 5, 6, 11 and 14 are assumed to be instantaneous (probability of an event = 1) since those step can occur completely (Hei et al., 1998; Chen et al., 2001a; Chen et al., 2001b). The diffusion of adsorbed species is also considered. However, only the diffusion of adsorbed $CH_4$, O, CO and H is included because of their relatively rapid diffusion rates (Ren et al., 2008; Chen et al., 2001b).

The adsorption constants ($k_i$) for methane and oxygen in steps 1 and 7 depend on the reactant pressures in the gas phase estimated by Eq. 1, where $k_i^0$ is adsorption coefficient and $y_i$ is the feed concentration and $P$ is the pressure of the reactants.

$$k_i = k_i^0 y_i P \qquad (1)$$

The reaction constant ($k_i$) in steps 2, 8, 9, 10, 12, 13, 15, 16, 17 and 18 can be calculated by the Arrhenius equation (Eq. 2), where $A_i$ is the pre-exponential factor, $Ea_i$ is the activation energy, $R$ is the gas constant and $T$ is the absolute temperature.

$$k_i = A_i \exp(-Ea_i / RT) \qquad (2)$$

The diffusion constant ($k_{diff}$) of adsorbed $CH_4$, O, CO and H can be calculated by Eq. 3, where $D_{0i}$ is the pre-exponential factor for the diffusion process, $a$ is the cell parameter of the nickel surface with a value of $2.48 \times 10^{-8}$ cm (Ren and Guoa, 2008) and $Q_i$ is the activation energy for the diffusion process. Because the kinetic parameters for the diffusion of adsorbed $CH_4$, CO and H are not precisely known, we use the kinetic parameters of adsorbed O for all diffusion species; $D_{0i} = 5.45 \times 10^{-3}$ cm$^2$ s$^{-1}$, $Q_i = 37.9$ kcal/mol (Nam et al., 2013; Barlow and Grundy, 1969). The effect of diffusion on the overall reaction is small, so knowing the precise diffusion rate is not crucial.

$$k_{diff} = \frac{D_{0i}}{a^2} \exp(-Q_i / RT) \qquad (3)$$

The diffusion number ($D_N$) is defined to represent the relative rate of diffusion and reaction (Eq. 4). This indicates that adsorbed species diffuse $D_N$ times faster than the reaction of the LH step (step 9) (Ren and Guoa, 2008).

$$D_N = \frac{k_{diff}}{k_{LH}} \qquad (4)$$

**2.2 Simulation procedure**

The KMC algorithm developed in this work is similar to that of Cortés et al. (2006, 2014). In the simulation, the surface of nickel metal is represented by a two-dimensional square lattice of $LxL$ sites. Periodic boundary conditions are applied to avoid edge effects. The surface is in contact with an infinite reservoir of methane and oxygen gas molecules with fixed feed concentrations $y_{CH4}$ and $y_{O2}$. Computing time is measured in MC cycles, defined as $L^2$ attempts of the reaction events listed in (c) below. Our KMC algorithm consists of the following steps:

**(a)** Choose one site from the surface randomly.

**(b)** Perform the instantaneous event (steps 3, 4, 5, 6, 11 and 14) if possible. On the site selected in step (a), we check neighboring sites for instantaneous event, whether there is a possible event or not. If the event is possible, the corresponding step is carried out and the surface is changed according to the event. This procedure is repeated if there is further possible event on that site. In the case of more than one possible instantaneous event, such as steps 11 and 14, one of them is chosen randomly and carried out. In the case of chain reaction in neighboring sites, all instantaneous events should be carried out immediately. However, to simplify the program we did not consider the chain reaction in neighboring sites in this study. Because these chain reactions are relatively rare, this assumption should not have too great an effect on the behavior.

**(c)** Choose a reaction event $i$ from the mechanism steps ($i$ = 1, 2, 7, 8, 9, 10, 12, 13, 15, 16, 17 and 18; excluding the instantaneous steps) according to the probability of an event $i$ ($p_i$) defined by Eq. 5. This procedure is known in different sources variously as the n-fold way or Bortz-Kalos-Lebowitz (BKL) or Gillespie algorithm (Bortz et al., 1975; Gillespie, 1976).

$$p_i = \frac{k_i}{\Sigma k_i} \tag{5}$$

**(d)** Perform the reaction event $i$ selected in step (c) according to the following processes:

*Adsorption*

• If the adsorption of $CH_4$ (step 1) is selected and the site selected in step (a) is empty, the event is successful, the adsorption of $CH_4$ is carried out and a particle of $CH_4^*$ is placed in the site. If the site is occupied, the attempt is terminated.

• If the dissociative adsorption of $O_2$ (step 7) is selected and the site selected in step (a) is empty, a neighboring site is then chosen randomly next to the first site. If the latter site is empty, the event is successful, the adsorption of $O_2$ is carried out and a particle of $O^*$ is placed in each of the two previous sites. If either site is occupied, the attempt is terminated.

*Desorption*

• If the desorption of $CH_4$ (step 2) is selected and the site selected in step (a) is occupied by a $CH_4^*$ particle, the event is successful, the desorption of $CH_4$ is carried out, the site is empty and a molecule of $CH_4$ leaves the surface. If the site is not occupied by a $CH_4^*$ particle, the attempt is terminated.

• If the desorption of CO (step 10) is selected, the procedure is completely analogous to the desorption of $CH_4$ (step 2).

• If the associative desorption of $O_2$ (step 8) is selected and the site selected in step (a) is occupied by an $O^*$ particle, a neighboring site is then chosen randomly next to the first site. If the latter site is occupied by the $O^*$ particle, the event is successful, the desorption of $O_2$ is carried out, the two sites are empty and a molecule of $O_2$ leaves the surface. If either site is not occupied by an $O^*$ particle, the attempt is terminated.

*Surface Reaction*

• If a surface reaction event (steps 9, 13, 16 or 18) is selected and the site selected in step (a) is occupied by a particle corresponding to one of the reactants, a neighboring site is then chosen randomly next to the first site. If the latter site is occupied by the other species of the same reaction, the event is successful, the corresponding reaction is carried out, one particle is replaced by a product particle and the other site is empty. If both sites are not occupied by the appropriate reactants, the attempt is terminated.

*Surface reaction and desorption*

- If the surface reaction and desorption event (steps 12 or 17) is selected, the procedure is analogous to a surface reaction event (step 9, 13, 16, 18) but a molecule of product leaves the surface and the site becomes empty.

*Formation of nickel oxide*

- If the formation of nickel oxide (step 15) is selected and the site selected in step (a) is occupied by an O* particle, the event is successful, the formation of nickel oxide is carried out and the site is replaced by Ox. If the site is not occupied by the O* particle, the attempt is terminated.

**(e)** Perform the diffusion event in proportional to diffusion number ($D_N$). One site from the surface is randomly chosen. If the site is occupied by a diffusion particle ($CH_4^*$, $O^*$, $CO^*$ or $H^*$) and a randomly chosen neighboring site is empty, the diffusion is successful, the diffusion particle moves to the chosen site. This procedure is repeated an average of $D_N$ times.

**(f)** Update time from $t$ to $t + \Delta t$ using Eq. 6, where $r$ is a uniformly distributed random number between 0 and 1, $L$ is lattice length and $\Sigma k_i$ is the sum of all reaction constants (excluding the instantaneous events).

$$\Delta t = \frac{-\ln r}{L^2 \Sigma k_i} \quad (6)$$

**(g)** If the stop condition is fulfilled, then the simulation is stopped. If not, repeat the algorithm.

In this simulation, runs up to 50,000 MC cycles were carried out. In order to get the average value, the initial 10,000 MC cycles were neglected to avoid non-equilibrium behavior, and the production rate ($R_i$), fractional coverage ($\theta_i$) and selectivity ($S_i$) were computed by taking averages over the subsequent 40,000 MC cycles. The production rates of produced $H_2$, CO, $CO_2$, $H_2O$ are determined by the numbers of product molecules per lattice site in a unit time. The selectivities of $H_2$ and CO are calculated by Eq. 7 and Eq. 8, respectively.

$$S_{H_2} = \frac{R_{H_2}}{R_{H_2} + R_{H_2O}} \quad (7)$$

$$S_{CO} = \frac{R_{CO}}{R_{CO} + R_{CO_2}} \quad (8)$$

The simulations were started with a 100% metallic nickel surface (empty surface) consisting of 128x128 sites, the feed concentration of methane and oxygen selected as 0.667 and 0.333, respectively, the pressure of the reactants was chosen as 6 kPa, and the temperature chosen as 873 K, corresponding to the experiment condition in (Tsipouriari and Verykios, 1998). The number of neighboring sites corresponding to lattice coordination number was set to four and the surface was considered free of impurities. Then, the lattice size, initial state configuration, temperature, feed concentration, lattice coordination number, diffusion number, and fraction of impurity were adjusted and the results of the different systems were evaluated.

**3. Results and discussion**

It is important to check the effect of the lattice size. Increasing of the lattice size from 32x32 to 256x256 sites causes only a slight difference in the production rates and fractional coverages, and the overall qualitative nature is not affected. To save the computing time, a lattice size of 128x128 sites was used in our simulations. At this size, there is less than 2% difference in critical values compared to 256x256 sites.

Simulations were performed with different initial configurations of the catalyst's surface, from 100% metallic (reduced) nickel surface to the oxidized nickel surface covered with 95% Ox. In all cases, the same results were obtained at steady state with nearly the same production rates and fractional coverages. However, for the oxidized nickel surface, it was observed that the simulation takes a longer time to reach steady state, especially when the surface is covered with a large amount of Ox. This is because the reaction with Ox (steps 16-18) exhibits much lower activity than most other reactions.

### 3.1. Effect of feed concentration

Fig. 1 presents the production rates as a function of the feed concentration of methane ($y_{CH4}$) in the gas phase at 873 K. It can be observed that the reaction occurs only for $y_1 < y_{CH4} < y_2$, where $y_1 = 0.47 \pm 0.01$ and $y_2 = 0.71 \pm 0.01$. Outside of this range, the surface is saturated (poisoned) with various species and the reaction cannot take place. The relation of fractional coverages and $y_{CH4}$ is also shown in Fig. 2. According to the result, Ox, C, H* and O* are found in a high portion on the surface while the other species are present at a much lower amount.

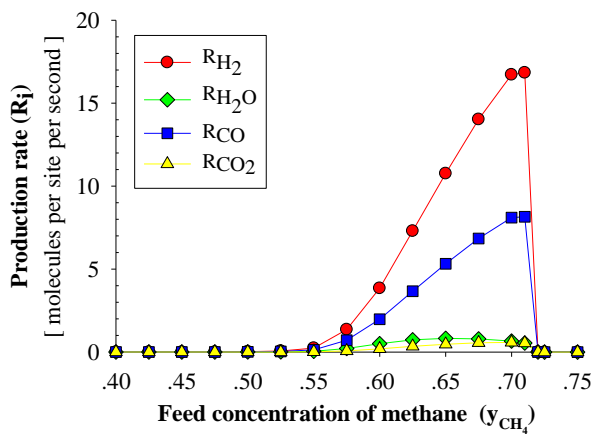 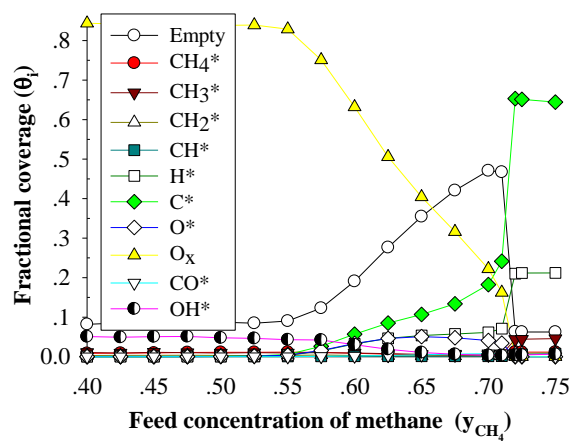

**Fig. 1.** Production rates as functions of $y_{CH4}$; $T = 873$ K, $P_{CH4+O2} = 6$ kPa, $Z = 4$, $D_N = 4.02$, $I = 0$.

**Fig. 2.** Fractional coverages as functions of $y_{CH4}$; $T = 873$ K, $P_{CH4+O2} = 6$ kPa, $Z = 4$, $D_N = 4.02$, $I = 0$.

As can be seen in Figs. 1 and 2, when $y_{CH4}$ is increased toward 0.71, the production rates of H$_2$ and CO continuously increase. However, when 0.71 is reached, C* suddenly grows and covers much of the lattice. The surface becomes saturated mainly with C*, H* and CH$_3$*, and the reaction stops. It can be said that the catalyst is deactivated due to carbon accumulation (poisoning by C*) on a nickel catalyst, as observed in experiments by Qiangu et al. (2000). This deactivation occurs abruptly, with discontinuities of the production rates and fractional coverages, implying that this is a first-order kinetic phase transition (Ziff et al., 1986). Similarly, when $y_{CH4}$ is lowered to 0.47, the surface becomes saturated mainly with Ox and OH*. It corresponds to the non-reactive state in which the surface is oxidized (Lashina et al., 2012). However, this transition is continuous; therefore, it is second-order. In an experiment, the poisoning by Ox may be considered as a reversible process because the possibility of the adsorption of reactants on the Ox surface (Eley–Rideal mechanism) can occur, and when the atmosphere is returned to a condition that allows steady-state operation, the reactive state returns.

For combustion products, the production rate of CO$_2$ has a similar trend as that of H$_2$ and CO but the production rate of H$_2$O has a different trend. It increases up to a maximum value at $y_{CH4} = 0.65$ and then decreases for increasing $y_{CH4}$ (Fig. 1). This is because the concentration of OH* decreases and the O* preferentially reacts with the C* to make CO. Note that while the production rates of $H_2$ and CO drop precipitously at $y_2$, the production rate of $H_2O$ goes continuously to zero.

Furthermore, it is noted that the maximum $H_2$ and CO production rates (Fig. 1) and the intersection of the C* and Ox fractional coverages (Fig. 2) are at the same $y_{CH4}$. This suggests that the reaction between C* and Ox (Step 16) is the rate-controlling step at this condition, and therefore, the maximum production rates are achieved when the coverages of C* and Ox are equal.

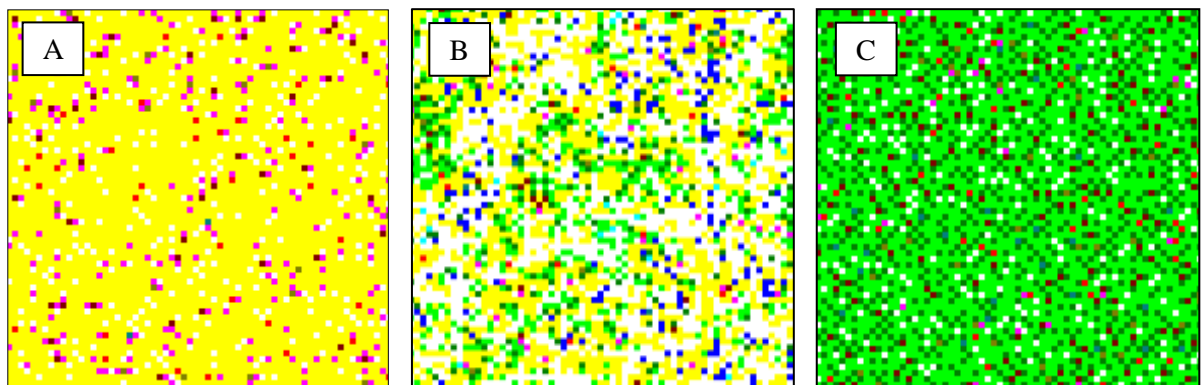

**Fig. 3.** Snapshots of ¼ of the surface at (A) $y_{CH4} = 0.40$ (B) $y_{CH4} = 0.67$ (C) $y_{CH4} = 0.75$;
$T = 873$ K, $P_{CH4+O2} = 6$ kPa, $Z = 4$, $D_N = 4.02$, $I = 0$.
(■ = Ox, □ = Empty, ■ = H*, ■ = C*, ■ = O*, ■ = CH$_4$*, ■ = CH$_3$*, ■ = CH$_2$*, ■ = CH*, ■ = CO*, ■ = OH*)

Fig. 3A illustrates a snapshot of the non-reactive state in which the surface is poisoned with Ox. Fig. 3B shows a snapshot of the steady reactive state at $y_{CH4} = 0.67$. At this condition, most of the surface is seen to be covered by C*, Ox and O*. This shows the progress of CO* formation (Steps 9 and 16) in the surface when C* reacts with a neighboring Ox or O* to form CO* and then the CO* is desorbed, leaving one empty site. These empty sites subsequently allow the new incoming reactants to adsorb resulting in the reaction to proceed. The snapshot corresponding to the carbon accumulation at $y_{CH4} = 0.75$ is shown in Fig. 3C.

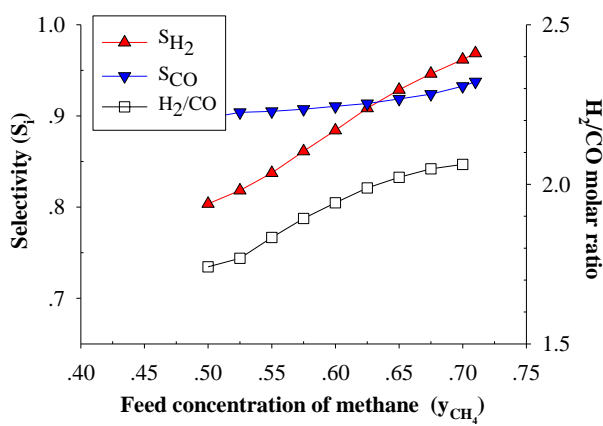

**Fig. 4.** Selectivities and $H_2$/CO molar ratio as functions of $y_{CH4}$;
$T = 873$ K, $P_{CH4+O2} = 6$ kPa, $Z = 4$, $D_N = 4.02$, $I = 0$.

The $H_2$ and CO selectivities and $H_2$/CO molar ratio are plotted as a function of $y_{CH4}$ in Fig. 4. Increasing methane in the feed results in the increase of $H_2$ and CO selectivities, as experimentally observed by Wang et al. (2009). In this model, the CPOx of methane follows the direct mechanism whereby methane directly converts to $H_2$ and CO, and $H_2O$ and $CO_2$ are formed by subsequent oxidation of $H_2$ and CO. The decrease of the oxygen in the feed leads to less oxidation of $H_2$ and CO. Therefore, the $H_2$ and CO selectivities increase with increasing methane feed concentration (Eriksson et al., 2007). The $H_2$/CO molar ratio in the products also increases with the increasing $y_{CH4}$. This suggests that the increase of the methane in the feed gives more $H_2$ than CO. Additionally, the $H_2$/CO molar ratio is around two for the entire range of feed concentrations, as expected from the stoichiometry.

### 3.2. Effect of reaction temperature

The production rates, fractional coverages, selectivities and $H_2$/CO molar ratio as a function of the reaction temperature are respectively shown in Figs. 5 to 7, when $y_{CH4} = 0.67$. At temperatures below 850 K, the catalyst is deactivated as a result of carbon accumulation on the nickel surface. The fractional coverages as depicted in Fig. 6 show that the coverage of C* decreases with increasing temperature, thus suppressing the carbon accumulation, and a steady reactive state is reached at temperatures higher than 850 K.

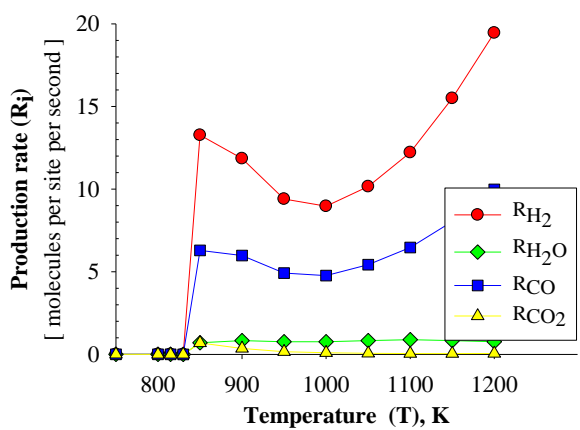 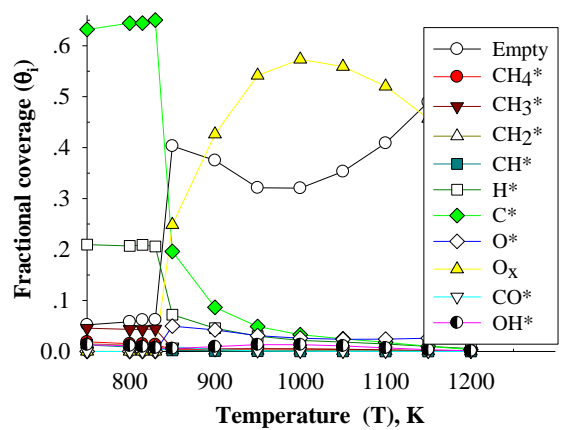

**Fig. 5.** Production rates as functions of $T$; $y_{CH4} = 0.67$, $P_{CH4+O2} = 6$ kPa, $Z = 4$, $D_N = 4.02$, $I = 0$.

**Fig. 6.** Fractional coverages as functions of $T$; $y_{CH4} = 0.67$, $P_{CH4+O2} = 6$ kPa, $Z = 4$, $D_N = 4.02$, $I = 0$.

For the reactive regions, the production rates of $H_2$ and CO decrease with increasing temperature and reach minimum values at a temperature around 1000 K (Fig. 5). Above 1000 K, the rate increases continuously as the temperature increases. This behavior is closely related to the fractional coverages as shown in Fig. 6.

When the temperature is below 1000 K, the reaction between C* and Ox (step 16) is the controlling step. Since the high fractional coverages of C* and Ox are dominant, the formation of CO is mainly due to the reaction between C* and Ox. The Ox formation increases with increasing temperature, causing a lower number of empty sites. As a result, the production rates of $H_2$ and CO are suppressed.

However, when the temperature is over 1000 K, the reaction between C* and O* (step 9) becomes the controlling step, because the fractional coverage of C* is lower than that of O* and the C* prefers to react with O* rather than Ox. Furthermore, past 1000 K, the Ox coverage decreases substantially. Therefore, the formation of CO is mainly due to the reaction between C* and O*. Because the reaction rate increases more rapidly than the Ox formation rate when temperature increases, the Ox coverage decreases with increasing temperature, causing a higher number of empty sites. As a result, the production rates of $H_2$ and CO increase as the temperature increases.

Notice in Figs. 5 and 6, the $H_2$ and CO production rates have a similar trend as the fraction of empty sites. The result indicates that the surface state of the catalyst plays a very important role in the kinetic behavior. The Ox, which is the oxidized surface of nickel, is much less reactive than the reduced or empty surface (Li et al., 2000). Apart from the main product, the production rate of $H_2O$ remains unchanged when the temperature goes up. The production rate of $CO_2$ falls rapidly with increasing temperature, and the selectivity of CO over $CO_2$ increases up to 0.99 at temperatures above 1075K.

As shown in Fig. 7, the CO selectivity increases significantly with an increase in temperature, due to the preference to undergo the total oxidation at lower temperatures resulting in $CO_2$ as the product (steps 12 and 17). The $H_2$ selectivity also increases at temperatures higher than 1000K. This appears to be related to the change in the production rate curves at 1000 K (Fig. 5). These trends are similar to the experimental results reported by Larimi et al. (2012). They found that higher temperature favors the direct CO and $H_2$ formation, and the CO selectivity becomes more pronounced than $H_2$ selectivity. Fig. 7 also presents the $H_2$/CO molar ratio as a function of temperature. It suggests that the increase of temperature leads to more CO than $H_2$ product.

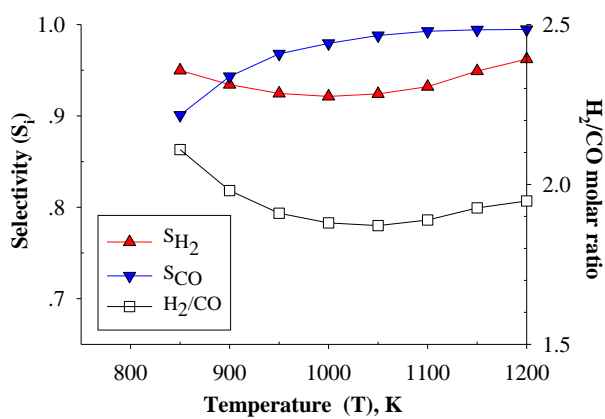

**Fig. 7.** Selectivities and $H_2$/CO molar ratio as functions of $T$; $y_{CH4} = 0.67$, $P_{CH4+O2} = 6$ kPa, $Z = 4$, $D_N = 4.02$, $I = 0$.

### 3.3. Effect of lattice coordination number

In this section, the effect of the lattice coordination number is studied. One way of changing the lattice coordination is to allow diagonal interactions on a square lattice. In addition to the four nearest neighbours (nn) sites, the diagonal sites or next nearest neighbours (nnn) sites are also included, increasing the coordination number ($Z$) of the site from four to eight. It is assumed that the nn and nnn sites are equivalent. This effect has been reported for the oxidation of CO and the reduction reaction of NO by CO and it seems to be a realistic way of modelling the surface (Cortés et al., 1998; Cortés, 1999; Valencia, 2000).

In the case of eight-site coordination, the steady reactive state appears when $0.42\pm0.01 < y_{CH4} < 0.71\pm0.01$, as shown in Figs. 8 and 9. The width of the steady reactive state is larger than in the four-site coordination case, where $0.47\pm0.01 < y_{CH4} < 0.71\pm0.01$. This is consistent with the results of Casties et al. (1998) who observe an enlargement of the steady reactive state when they allow diagonal interactions. It is found that $y_2$, the transition at high concentration of methane, remains constant at 0.71. However, it is seen that $y_1$, the transition at low concentration of methane, decreases from 0.47 to 0.42 when increasing the coordination number. The reason might be that an increase in the possibility of O* reactions in diagonal interactions (step 9, 12 and 13) suppresses the possibility of Ox formation from O* (step 15), requiring a greater feed concentration of oxygen to poison the surface with Ox. Figs. 8 and 9 also shows the character of the phase transition in the eight-site coordination case, which appears to be continuous at $y_2$, in contrast to the discontinuous behavior in the four-site coordination case (Figs. 1 and 2). This is also in contrast to the ZGB model, which adding diagonal interactions do not produce a continuous transition (Cortés, 1999). The value of the maximum production rates of products is not much different in both cases. However, if the eight-site coordination is considered, the Ox coverage in the Ox-poisoned region increases slightly and C* coverage in C*-poisoned regions increases significantly.

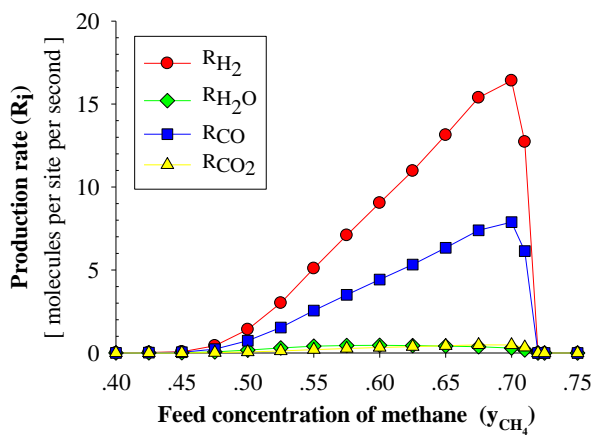 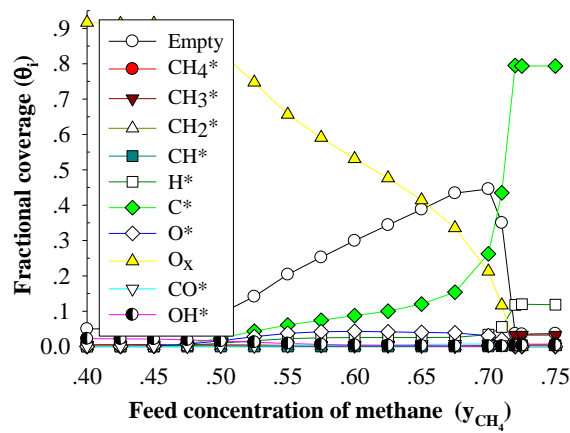

**Fig. 8.** The same as Fig. 1, but for $Z = 8$.  **Fig. 9.** The same as Fig. 2, but for $Z = 8$.

### 3.4. Effect of diffusion of adsorbed species

To understand the effects of diffusion, the production rates of $H_2$ and CO as a function of $y_{CH4}$ are presented at different diffusion numbers: $D_N = 0$, 0.1, 1, 4.02 (the number used in the other simulations) and 10. From Figs. 10 and 11, two important observations can be made. First, the maximum production rates of $H_2$ and CO increase moderately with the increase of $D_N$. It is generally expected that diffusion increases the production rates, since the mobility of the adsorbed species facilitates the number of encounters between the reactants. Second, it is found that, with the increase of diffusion rate, the transition points ($y_1$ and $y_2$) increase and the width of the steady reactive state increases slightly. Table 2 summarizes the transition points and the widths of steady reactive states for different values of $D_N$. It can be seen that the result is almost the same in the case of $D_N = 0$ and $D_N = 0.1$ and the increasing of $D_N$ from 0 to 10 causes only a slight difference in the transition points and the widths of steady reactive states. This result indicates that the diffusion rate plays a little role in this reaction system. The reason is that only adsorbed $CH_4$, O, H and CO are allowed to diffuse and these adsorbed species have low fractional coverages on the surface ($\theta_i < 0.1$). The fractional coverages of each species are similar to that obtained in the absence of diffusion. The only notable difference is in Ox coverage, which increases as $D_N$ grows, as shown in Fig. 12. When Ox coverage increases, it requires a greater feed concentration of methane to poison the surface with C* and requires a lesser feed concentration of oxygen to poison the surface with Ox. This makes the transition points moved towards higher values of $y_{CH4}$ with the increase of $D_N$.

The effect of changing which species are allowed to diffuse was also studied and it is observed that diffusion of $CH_4$* and CO* has little effect on the reaction due to their extremely low fractional coverages ($\theta_i < 0.01$). In addition, the diffusion of O* has more impact on the production rates than H*. Inclusion of the diffusion of O* produces the largest increase in the production rates and width of the steady reactive state. However, the selectivity of $H_2$ decreases significantly when the diffusion of O* is introduced, which promotes $H_2O$ formation.

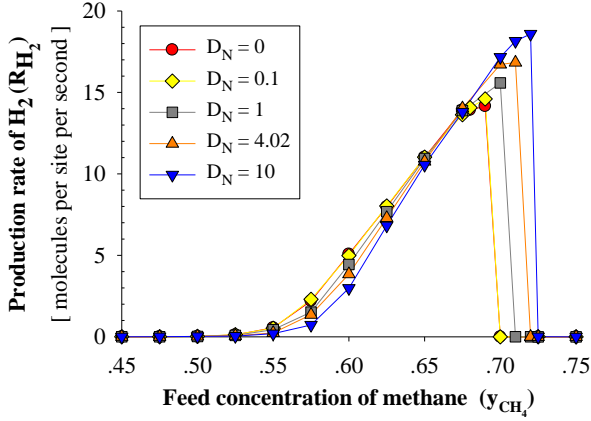

**Fig. 10.** Production rate of H$_2$ as functions of $y_{CH4}$ for different values of $D_N$; $T$ = 873 K, $P_{CH4+O2}$ = 6 kPa, $Z$ = 4, $I$ = 0.

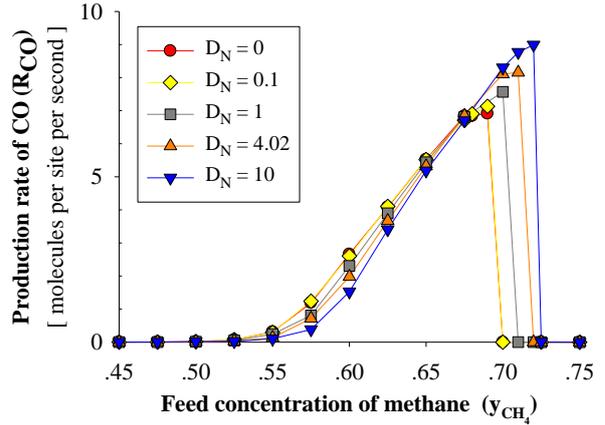

**Fig. 11.** Production rate of CO as functions of $y_{CH4}$ for different values of $D_N$; $T$ = 873 K, $P_{CH4+O2}$ = 6 kPa, $Z$ = 4, $I$ = 0.

**Table 2.** The transition points and the widths of steady reactive states for different value of $D_N$; $T$ = 873 K, $P_{CH4+O2}$ = 6 kPa, $Z$ = 4, $I$ = 0.

| $D_N$ | $y_1$ | $y_2$ | Widths ($y_1$-$y_2$) |
|---|---|---|---|
| 0 | 0.48 | 0.69 | 0.21 |
| 0.1 | 0.48 | 0.69 | 0.21 |
| 1 | 0.49 | 0.70 | 0.21 |
| 4.02 | 0.49 | 0.71 | 0.22 |
| 10 | 0.50 | 0.72 | 0.22 |

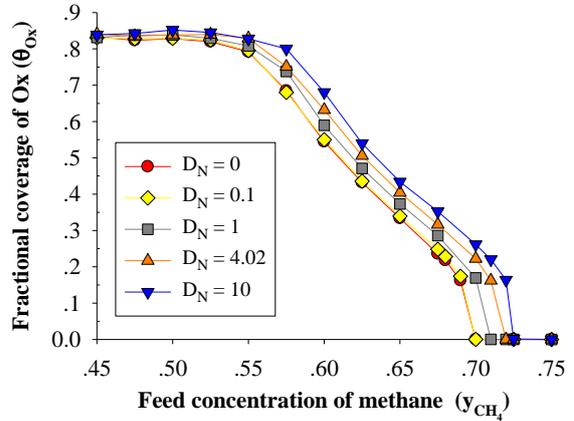

**Fig. 11.** Production rate of CO as functions of $y_{CH4}$ for different values of $D_N$; $T$ = 873 K, $P_{CH4+O2}$ = 6 kPa, $Z$ = 4, $I$ = 0.

### 3.5. Effect of inactive impurities on a surface

In experimental systems, the surface can be contaminated with impurities such as sulfur. The impurities can adsorb on the surface, where they do not react with the other adsorbed species. These impurities play an important role in the reaction process. Bitsch-Larsen et al. (2008) experimentally investigated the effect of sulfur in CPOx and found that the presence of small amounts of sulfur have a negative impact on the reaction significantly, decreasing the CH$_4$ conversion and H$_2$ selectivity. In this work, the impurity fraction ($I$) is defined as the fraction of inactive impurities which are on the surface, and these impurities are randomly distributed on the surface. The sites adsorbed by these impurities cannot take part in reaction or diffusion processes.

The behavior of a system with impurities is shown in Figs. 13 and 14, where we plot the production rates of H$_2$ and CO versus $y_{CH4}$ for different value of impurities fraction: $I$ = 0, 0.1, 0.2 and 0.3. It is observed that with the increase of impurities, the production rates of H$_2$ and CO significantly decrease. Furthermore, the transition points move toward a lower value of $y_{CH4}$ with the increase of impurities, as shown in Table 3. The reason is that when impurities increase, the number of neighboring free sites decreases. Therefore, oxygen, which requires two adjacent sites to adsorb, has a lower possibility of adsorbing on the surface. Methane prefers to adsorb rather than oxygen and the transition points shift to lower values of $y_{CH4}$. However, the widths of the steady reactive state shrink with an increase in impurity fraction, as shown in Table 3, and the reaction stops when the impurity fraction exceeds 0.53. The interesting observation in this system is the conversion of the first-order discontinuous phase transition into a second-order continuous phase transition when impurities are added. As seen in Figs. 13 and 14, the discontinuous transition is softened with the addition of impurities and further converts into a continuous transition. This phenomenon has also been observed previously by Hoenicke and Figueiredo (2000) and Lorenz et al. (2002) in the CO-$O_2$ reaction and Ahmad et al. (2007) in the NO-CO reaction. It should be noted that the effect of impurities in CPOx has not been studied before. This conversion of a discontinuous phase transition to a continuous one is important because in the discontinuous phase transition, the fluctuations in operating conditions may lead the system into the C* poisoning region and the system cannot come back to the reactive region. However, for continuous phase transitions, the reaction is sustained and the system can re-enter into the steady reactive state if the atmosphere is returned (Ahmad et al., 2007). Fig. 15 shows the fractional coverages as a function of $y_{CH4}$ at $I = 0.3$. Comparing with the case of no impurities (Fig. 2), it can be seen that, in reactive region, the coverages of O* and Ox decrease while the coverages of C* and H* increase. This implies that methane prefers to adsorb rather than oxygen if there are impurities on the surface.

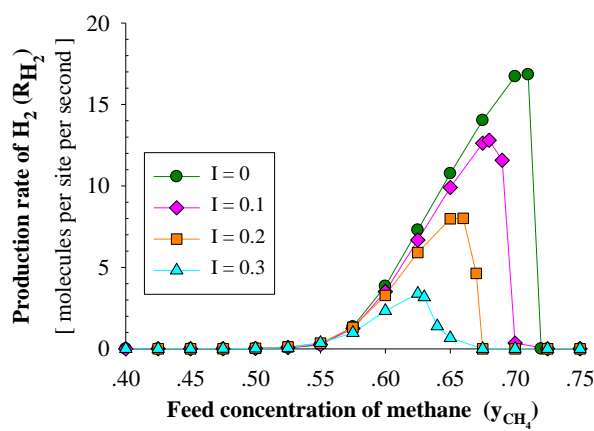

**Fig. 13.** Production rate of $H_2$ as functions of $y_{CH4}$ for different values of $I$; $T = 873$ K, $P_{CH4+O2} = 6$ kPa, $Z = 4$, $D_N = 4.02$.

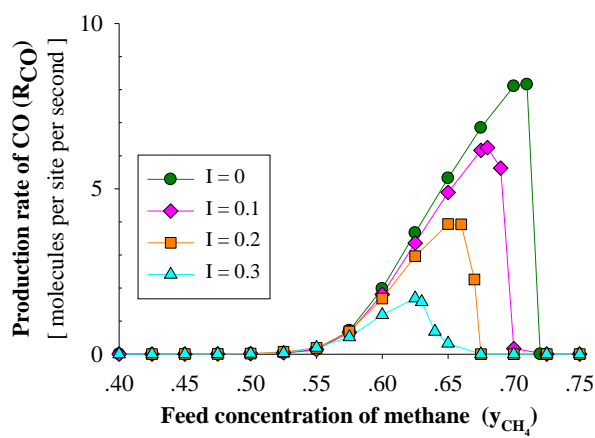

**Fig. 14.** Production rate of CO as functions of $y_{CH4}$ for different values of $I$; $T = 873$ K, $P_{CH4+O2} = 6$ kPa, $Z = 4$, $D_N = 4.02$.

**Table 3.** The transition points and the widths of steady reactive states for different value of $I$; $T = 873$ K, $P_{CH4+O2} = 6$ kPa, $Z = 4$, $D_N = 4.02$.

| $I$ | $y_1$ | $y_2$ | Widths ($y_1$-$y_2$) |
|---|---|---|---|
| 0 | 0.49 | 0.71 | 0.22 |
| 0.1 | 0.49 | 0.69 | 0.20 |
| 0.2 | 0.47 | 0.67 | 0.20 |
| 0.3 | 0.47 | 0.66 | 0.19 |

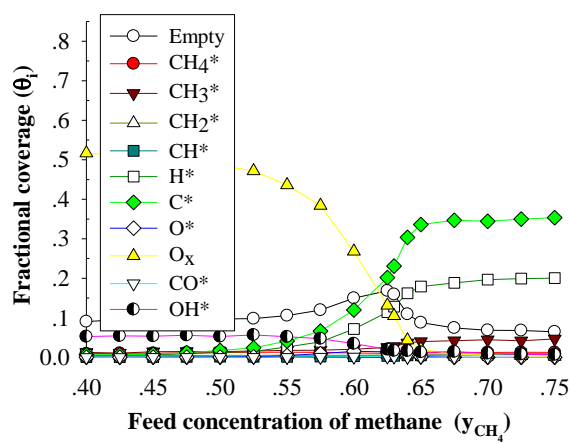

**Fig. 15.** Fractional coverages as functions of $y_{CH4}$ at $I=0.3$; $T = 873$ K, $P_{CH4+O2} = 6$ kPa, $Z = 4$, $D_N = 4.02$.

## 4. Conclusions

The partial oxidation of methane over a nickel catalyst is simulated by KMC method. According to the simulation results, the reaction occurs only in a certain range of methane feed concentration and reaction temperature. Outside of this range, the surface is poisoned mainly with nickel oxide or carbon. It is shown that both first- and second-order phase transitions occur. This behavior is similar to the ZGB model for the oxidation reaction of CO because both these models essentially deal with the monomer-dimer reaction. It is observed that with the increase of reaction temperature, the controlling step can be changed from the reduction of Ox by adsorbed C (step 16) to the reaction between adsorbed C and O (step 9). Our results agree with experiments concerning the feed concentration and temperature effects on the selectivities and $H_2$/CO molar ratio. However, experiments showing the fractional coverages to compare with do not seem to be available. The influence of lattice coordination number is investigated and the result suggests that the width of the steady reactive state is significantly larger if diagonal interactions are allowed. We also find that the first-order phase transition smooths out with eight-site coordination. An increase of the diffusion rate slightly increases the maximum production rates and width of the steady reactive state. Among all adsorbed species, O and H play the most important roles in the diffusion process although the effect is small. When the surface is contaminated with impurities, the production rates are greatly reduced and the reaction stops at a certain point. Furthermore, the discontinuous phase transition transforms into a continuous phase transition in the presence of impurities. Future work on this reaction model would be to find out why diagonal interactions (eight-site coordination) lead to a continuous transition, and to study the behavior on supported catalysts.

## Acknowledgments

This work was supported by the Thailand Research Fund (TRF) under the TRF Senior Research Scholar (RTA578008).

## References

Ahmad, W., Kaleem, M., 2007. The effect of inactive impurities on a surface in NO-CO reaction : A Monte Carlo simulation. Applied Surface Science 253, 8447–8452.

Al-Sayari, S.A., 2013. Recent developments in the partial oxidation of methane to syngas. The Open Catalysis Journal 6, 17–28.

Barlow, R., Grundy, P.J., 1969. The determination of the diffusion constants of oxygen in nickel and α-iron by an internal oxidation method. Journal of Materials Science 4, 797–801.

Bitsch-Larsen, A., Degenstein, N.J., Schmidt, L.D., 2008. Effect of sulfur in catalytic partial oxidation of methane over Rh-Ce coated foam monoliths. Applied Catalysis B: Environmental 78, 364–370.

Bortz, A.B., Kalos, M. H., Lebowitz, J.L., 1975. A new algorithm for Monte Carlo simulation of Ising spin systems. Journal of Computational Physics 17, 10–18.

Chen, D., Lodeng, R., Anundskas, A., Olsvik, O., Holmen, A., 2001a. Deactivation during carbon dioxide reforming of methane over Ni catalyst: Microkinetic analysis. Chemical Engineering Science 56, 1371–1379.

Chen, D., Lodeng, R., Omdahl, K., Anundskas, A., Olsvik, O., and Holmen, A., 2001b. A model for reforming on Ni catalyst with carbon formation and deactivation. Studies in Surface Science and Catalysis 139, 93-100.

Cortés, J., 1999. Monte Carlo and mean field theory studies of the effect of the next nearest neighbour sites of a square lattice on the monomer-dimer surface reaction. Physical Chemistry Chemical Physics 1, 1577–1581.

Cortés, J., Valencia, E., 1998. Next nearest neighbors sites and the reactivity of the CO-NO surface reaction. Chemical Physics 229, 265–273.

Cortés, J., Valencia, E., Araya, P., 2006. Monte Carlo simulation studies of the catalytic combustion of methane. Catalysis Letters 112, 121–128.

Cortés, J., Valencia, E., Araya, P., 2014. Monte Carlo simulations in the preferential oxidation of carbon monoxide on a copper-ceria catalyst. Chemical Physics Letters 612, 97–100.

Eriksson, S., Rojas, S., Boutonnet, M., Fierro, J.L.G., 2007. Effect of Ce-doping on Rh/$ZrO_2$ catalysts for partial oxidation of methane. Applied Catalysis A: General 326, 8–16.

Evans, J.W., 1993. ZGB surface reaction model with high diffusion rates. Journal of Chemical Physics 98, 2463-2465.

Fichthorn, K., Gulari, E., Ziff, R.M., 1989. Self-sustained oscillations in a heterogeneous catalytic reaction: A Monte Carlo simulation. Chemical Engineering Science 44, 1403-1411.

Fichthorn, K.A., Weinberg, W.H, 1991. Theoretical foundations of dynamical Monte Carlo simulations. Journal of Chemical Physics 95, 1090-1096.

Gillespie, D.T., 1976. A general method for numerically simulating the stochastic time evolution of coupled chemical reactions. Journal of Computational Physics 22, 403–434.

Hei, M., Chen, H., Yi, J., Lin, Y., Lin, Y., Wei, G., Liao, D., 1998. $CO_2$-reforming of methane on transition metal surfaces. Surface Science 417, 82–96. doi:10.1016/S0039-6028(98)00663-3

Hoenicke, G.L., Figueiredo, W., 2000. ZGB model with random distribution of inert sites. Physical Review E 62, 6213-6223.

Hu, Y.H., Ruckenstein, E., 1996. Transient kinetic studies of partial oxidation of $CH_4$. Journal of Catalysis 158, 260–266.

Larimi, A.S., Alavi, S.M., 2012. Partial oxidation of methane over Ni/$CeZrO_2$ mixed oxide solid solution catalysts. International journal of Chemical Engineering and Applications 3, 6–9.

Lashina, E. A., Kaichev, V. V., Chumakova, N. A., Ustyugov, V. V., Chumakov, G. A., Bukhtiyarov, V.I., 2012. Mathematical simulation of self-oscillations in methane oxidation on nickel: an isothermal model. Kinetics and Catalysis 53, 374–383.

Li, C., Yu, C., Shen, S., 2000. Role of the surface state of Ni/$Al_2O_3$ in partial oxidation of $CH_4$. Catalyst Letters 67, 139–145.

Li, Y., Xiang, S., 2000. Micro-kinetic analysis and Monte Carlo simulation in methane partial oxidation into synthesis gas. Catalysis Today 61, 237–242.

Lorenz, C.D., Haghgooie, R., Kennebrew, C., Ziff, R.M., 2002. The effects of surface defects in a catalysis model. Surface Science 517, 75-86.

Loscar, E.S., Albano, E.V., 2003. Critical Behaviour of Irreversible Reaction Systems. Reports on Progress in Physics 66, 1343-1382.

Machado, E., Buendía, G.M, Rikvold, P.A., Ziff, R.M., 2005. Response of a catalytic reaction to periodic variation of the CO pressure: Increased $CO_2$ production and dynamic phase transition. Physical Review E 71, 016120.

Nam, H.O., Hwang, I.S., Lee, K.H., Kim, J.H., 2013. A first-principles study of the diffusion of atomic oxygen in nickel. Corrosion Science 75, 248–255.

Qiangu, Y., Tinghua, W., Jitao, L.,Chunrong, L., Weizeng, W., Lefu, Y., Huilin, W., 2000. Mechanism study of carbon deposition on a Ni/$Ai_2O_3$ catalyst during partial oxidation of methane to syngas. Journal of Natural Gas Chemistry 9, 89-92.

Ren, X. and Guoa, X., 2008. Monte Carlo simulation of the oscillatory behavior in partial oxidation of methane on nickel catalyst under nonisothermal conditions. Surface Science 603, 606-610.

Ren, X., Lia, H., and Guoa, X., 2008. Monte Carlo simulation of the oscillatory behavior in partial oxidation of methane on nickel catalyst. Surface Science 602, 300-306.

Shen, S., Li, C., and Yu, C., 1998. Mechanistic study of partial oxidation of methane to syngas over a Ni/Al$_2$O$_3$ catalyst. Studies in Surface Science and Catalysis 119, 765-770.

Smith, M.W., Shekhawat, D., 2011. Catalytic partial oxidation, Fuel cells: technologies for fuel processing. Elsevier.

Tomé, T., Dickman, R., 1993. Ziff-Gulari-Barshad model with CO desorption: An Ising-like nonequilibrium critical point. Physical Review E 47, 948-952.

Tsipouriari, V.A., Verykios, X.E., 1998. Kinetic study of the catalytic partial oxidation of methane to synthesis gas over Ni/La$_2$O$_3$ catalyst, Studies in Surface Science and Catalysis. Elsevier Masson SAS.

Valencia, E., 2000. Effect of the coordination of the superficial site in the monomer-dimer reaction on a disordered substrate. Surface Science 470, 109-115. doi: 10.1016/S0039-6028(00)00859-1

Wang, Y., Wang, W., Hong, X., Li, B., 2009. Zirconia promoted metallic nickel catalysts for the partial oxidation of methane to synthesis gas. Catalysis Communications 10, 940–944.

Ziff, R.M., Gulari, E., Barshad, Y., 1986. Kinetic phase transitions in an irreversible surface-reaction model. Physical Review Letters 56, 2553-2556.